%% file: paper.tex
\documentclass{juliacon}
\usepackage{amsmath}
\usepackage{amssymb}
\usepackage{cleveref}
\usepackage{microtype}
\begin{document}

% Misc typographical tweaks.
\setlength{\parindent}{10pt}
% \setmonofont{JuliaMono}[Extension = .ttf, Path = ./, UprightFont = *-Regular, ItalicFont = *-Italic]

\input{header}

\maketitle

\begin{abstract}
  Reliable numerical computations are central to scientific computing,
  but the floating-point arithmetic that enables large-scale models
  is error-prone.
  Numeric exceptions are a common occurrence and can propagate through
  code, leading to flawed results.
  This paper presents FlowFPX, a toolkit for systematically debugging
  floating-point exceptions by recording their flow,
  coalescing exception contexts, and fuzzing in select locations.
  These tools help scientists discover when exceptions happen
  and track down their origin, smoothing the way to a reliable codebase.
\end{abstract}

\newcommand{\code}[1]{\texttt{#1}}
\newcommand{\FlowFPX}{FlowFPX}
\newcommand{\GPUFPX}{GPU-FPX}
\newcommand{\FloatTracker}{FloatTracker}
\newcommand{\FT}{\FloatTracker}
\newcommand{\Fp}{Floating-point} % hyphen or no?
\newcommand{\fp}{floating-point} % hyphen or no?
\newcommand{\CSTG}{stack graph}
\newcommand{\CPP}{\code{C++}}
\newcommand{\Dendro}{\textsc{Dendro}}
\newcommand{\urlaccess}[2]{\url{#1}}
\newcommand{\Nan}{\code{NaN}}
\newcommand{\NaN}{\Nan}
\newcommand{\Inf}{\code{Inf}}
\newcommand{\zerowidth}[1]{\makebox[0pt][l]{#1}}
\newcommand{\zerocode}[1]{\zerowidth{\code{#1}}}
\newcommand{\genpropkill}{\emph{gen}-\emph{prop}-\emph{kill}}
\newcommand{\bigcheckmark}{\ding{51}}
\newcommand{\tblnext}{\(\Rightarrow\)}
\newcommand{\tblY}{\checkmark}
\newcommand{\tblN}{\scalebox{1.2}{$\times$}}

\section{Introduction}

Reliable numeric computations are central to high-performance computing,
machine learning, and scientific applications.
Yet the \fp{} arithmetic that powers these computations is fundamentally
unreliable~(\cref{s:background}).
Exceptional values, such as Not a Number (\Nan{}) and infinity (\Inf{}),
can and often do arise thanks to culprits such as roundoff error,
catastrophic cancellation, singular matrices, and vanishing
derivatives~\cite{sdjmrstp-pc-2022,ddghlllprr-correctness-2022,gllprt-correctness-2021,fpchecker-reports,llg-soap-2022,bllmg-xloop-2022}.
Developers are responsible for guarding against exceptions, but this task
is difficult because many operations can generate and propagate exceptions.
Worst of all, operations such as less-than (\code{<}) can kill an exceptional value,
leaving no trace of the problem.
There is little tool support to assist in exception debugging,
thus (unsurprisingly) a quick GitHub search reports over 4,000 open issues
related to \NaN{} or \Inf{} exceptions~\cite{github-issues}.

This paper introduces \FlowFPX{}, a toolkit for debugging
\fp{} exceptions~(\cref{s:flowfpx})
that has helped improve a variety of applications,
from ocean simulations to heat modes~(\cref{s:casestudies}):
\begin{itemize}
  \item
    The centerpiece of \FlowFPX{} is \FT{}, a dynamic analysis tool that
    selectively monitors for exceptions and fuzzes code for vulnerabilities.
  \item
    To visualize results, \FT{} adapts coalesced stack-trace graphs (CSTGs, or
    \CSTG{}s)~\cite{humphreySystematicDebuggingMethods2014}
    to summarize the program contexts that handled exceptional values.
  \item
    A companion tool \GPUFPX{}~\cite{llsflg-hpdc-2023} provides fine-grained insights for GPU kernels.
\end{itemize}

\section{Floating-Point Exception Primer}
\label{s:background}

\begin{figure}[t]\centering
  \includegraphics[trim=10 0 10 0,clip,width=0.95\columnwidth]{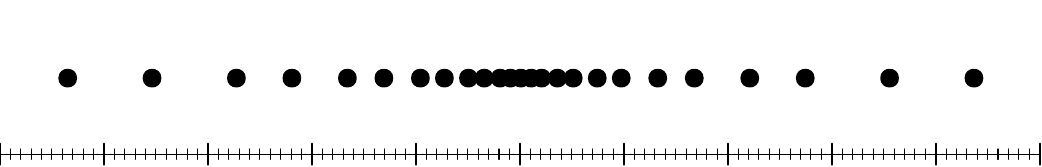}
  \caption{Floats are spread across the real number line}
  \label{f:real-vs-fp}
\end{figure}

\Fp{} numbers use a finite number of bits to represent a spectrum of points along the real number line~(\cref{f:real-vs-fp}).
The implementation strategy is essentially that followed in scientific notation.
A \fp{} number packs a sign bit, an exponent, and a fraction part (also called the ``significand'' or ``mantissa'') into a bitstring.
Typical strings are 64 or 32 bits long, but 16-bit and 8-bit formats are on the rise~\cite{klowerLowprecisionClimateComputing2021,fp8}.
This representation supports very small and very large numbers in a narrow range of bits:

\begin{lstlisting}[language = Julia]
julia> prevfloat(typemax(Float64))
1.7976931348623157e308
julia> nextfloat(typemin(Float64))
-1.7976931348623157e308
\end{lstlisting}

The flip side is that most real numbers fall into the gaps between \fp{} numbers and must be rounded, which introduces error
and can lead to surprising results.
For example, adding the tiny Planck constant to the large Avogadro number results
in Avogadro's number after rounding:

\begin{lstlisting}[language = Julia]
julia> planck = 6.62607015e-34
6.62607015e-34
julia> avogadro = 6.02214076e23
6.02214076e23
julia> avogadro + planck == avogadro
true
\end{lstlisting}

This is an extreme example, but many operations on \fp{} numbers closer in magnitude induce a loss of accuracy.
Refer to the literature for more details, e.g.,~\cite{knuthArtComputerProgramming1997,torontoPracticallyAccurateFloatingPoint2014,mullerHandbookFloatingPointArithmetic2018}.

\subsection{Exceptions and Exceptional Values}
\label{s:exnvalue}

The IEEE~754 \fp{} standard~\cite{IEEEStandardBinary1985}
defines exceptions and exceptional values as the outcome of
operations that have ``no single universally acceptable result''~\cite{p-draft-1997}.
For example, dividing by zero and exceeding the \code{Float64} range
both lead to exceptions:

% julia> 2 / 0
% Inf
\begin{lstlisting}[language = Julia]
julia> 0 / 0
NaN
julia> avogadro^avogadro
Inf
julia> log(0)
-Inf
julia> Inf + NaN
NaN
\end{lstlisting}

IEEE~754 defers the question of how to handle such exceptions to application code.
It is up to developers to watch for \Nan{}, \Inf{}, and subnormal numbers (underflow)
and implement an appropriate repair.
This is no easy task, however, because of the approximations inherent to
floating point.
Even an apparently-safe division could result in a \Nan{} if its
denominator gets truncated to zero.
Some \Nan{}s might be spurious, others might be fatal, but in any event
anticipating the various exceptions is a burden.
All too often, exceptional values go unhandled and flow through the code.

\subsection{Lifetime of an Unhandled Exceptions}
\label{s:to-kill-a-fp}

\begin{figure}[t]
  \includegraphics[width=\columnwidth]{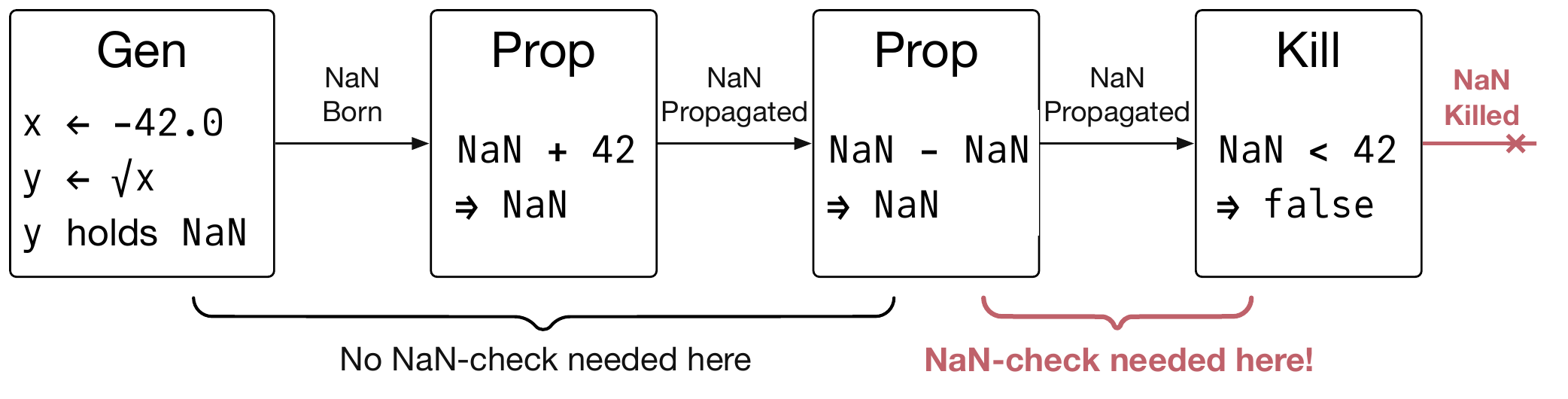}
  \caption{Gen, Prop, Kill: Lifetime of an exceptional value}
  \label{f:gpk}
\end{figure}

Unhandled exceptional values have a lifetime: they are born, or \emph{generated}, by some operation; they \emph{propagate} through other operations; and they either appear in the program output, go out of scope, or get \emph{killed} by a numeric operation.
\Cref{f:gpk} summarizes this \genpropkill{} process.
The gens and props are straightforward; see above for examples~(\cref{s:exnvalue}).
The kills often arise from numeric comparisons (\code{<}, \code{=}, etc.),
but exponents (\code{1{\string^}NaN}) and over-eager matrix
optimizations~\cite{ddghlllprr-correctness-2022} can kill exceptions as well.

To illustrate the perils of killed exceptions, consider the following two ways
of finding the maximum value in a list.
The first compares numbers with \code{<=}
while the second uses the built-in \code{max} function:

\begin{lstlisting}[language = Julia]
function max1(lst)
  max_seen = 0.0
  for x in lst
    # swap if x is not too small
    if ! (x <= max_seen)
      max_seen = x
    end
  end
  max_seen
end

function max2(lst)
  foldl(max, lst)
end
\end{lstlisting}

For lists with a \Nan{} inside, the functions can give different
results because \code{<=} kills \Nan{}s whereas \code{max}
propagates them:

\begin{lstlisting}[language = Julia]
julia> max1([1, 5, NaN, 4]) 
4.0
julia> max2([1, 5, NaN, 4]) 
NaN
\end{lstlisting}

Not only is the result from \code{max1} problematic for obscuring the fact that there was a \NaN{} in the list, the result is arguably wrong!
The result from \code{max2} at least shows that a \Nan{} was in the works, though in a realistic setting it may not be clear where the \Nan{} came from.
Both versions would thus benefit from tools that track exceptions across their lifetime.
FlowFPX can help.

\section{\FlowFPX{}}
\label{s:flowfpx}

\FlowFPX{} is a toolkit for tracking down \fp{} exceptions.
The primary tools in this paper are \FT{}, which records lifetimes
and enables fuzzing, and \CSTG{}s (more precisely, CSTGs), which visualize
the flow of exceptions.
\GPUFPX{} is a third component of \FlowFPX{} that tracks \fp{} exceptions inside GPU kernels~\cite{llsflg-hpdc-2023}.

\subsection{\FT{}}
\label{s:floattracker}

\FT{} tracks exceptional values across their \genpropkill{} lifetime
and can fuzz code for vulnerabilities by injecting a \Nan{} or \Inf{}
as the result of an operation.
\FT{} is implemented in Julia and is available on JuliaHub:

\noindent\begin{center}\noindent\!\!\url{https://juliahub.com/ui/Packages/FloatTracker/dBXig}\!\!
\end{center}

\subsubsection{Tracking Exceptional Values}
\label{s:trackingexceptionalvalues}

\begin{figure}[t]\centering
  \begin{tabular}{ccccc}
    Exn. Input? & \tblnext & Exn. Output? & = & Event       \\ \hline
    \tblN       & \tblnext &  \tblY{}     & = & {gen}  \\
    \tblY       & \tblnext &  \tblY{}     & = & {prop} \\
    \tblY       & \tblnext &  \tblN{}     & = & {kill} \\
  \end{tabular}
  \caption{How to classify operations that see exceptions}
  \label{fig:lifetime-class}
\end{figure}

\FT{} monitors \Nan{} and \Inf{} exceptions by overloading arithmetic
operations and logging key events.
When the input to an operation is exception-free but the output is
exceptional, the operation is a {gen} event~(\cref{fig:lifetime-class}).
When the input and output contain exceptions, the operation is a {prop}
event.
And when the input contains an exception but the output does not,
the operation is a {kill} event.
Put together, the log of all {gen}s, {prop}s, and {kill}s
sheds light of how various exceptions traveled across the program.
The logs also record the call context (stack trace) and the arguments to the operation as a starting point for debugging efforts.

\FT{} writes logs to three files: one for {gen} events, one for {prop}s, and one for {kill}s.
This way, discovering where a \Nan{} came from is a matter of sifting through the {gen} file.
Users can also lower the overhead of logging by turning it off for prop events.

The instrumentation works through custom \fp{} types: \code{TrackedFloat64},
\code{TrackedFloat32}, and \code{TrackedFloat16}.
Developers must opt in to \FT{} by wrapping numbers in a custom type.
From then on, tracking is automatic and extends transitively to all outputs
of tracked operations.
Multiplying a \code{Float64} value with a \code{TrackedFloat64}, for example, yields
a \code{TrackedFloat64} to continue the logging trail.

Crucially, the tracked version of an exception, say a \NaN{}, points
to the original exceptional value.
This enables techniques such as \NaN{}-packing, or otherwise
using the payload in creative ways.
Not all Julia operations currently preserve \NaN{}s
bit-for-bit,\footnote{\url{https://github.com/JuliaLang/julia/issues/48523}}
but \FT{} does its part.

\subsubsection{Fuzzing}

Operator overloading gives \FT{} a powerful way to fuzz code from the inside out.
Each overloaded function serves as a hook where \FT{} can decide whether to observe
the operation or step in, discard the original result, and return an exception instead.
Injecting faults in a random way~\cite{hamlet1994random}, also known as fuzzing,
is a useful way to discover vulnerabilities in a large codebase.
Demmel et~al. propose essentially the same idea for BLAS and
LAPACK~\cite{ddghlllprr-correctness-2022}.

\FT{} exposes several parameters to let developers control the fuzzer:

\begin{itemize}
\item \texttt{odds::Int64} inject if \texttt{rand(odds) == 1}.
\item \texttt{n\_inject::Int64} upper bound on the number of \Nan{}s to inject.
\item \texttt{active::Bool} fuzz only when set to true.
\item \texttt{functions::Array{String}} limit fuzzing to the dynamic extent of the listed functions.
\item \texttt{libraries::Array{String}} limit fuzzing to functions from the following libraries
\end{itemize}

When fuzzing reveals an error, the next step is to craft a regression test to
guide repairs.
\FT{} therefore records the sequence of injections that it makes during a fuzzing
run and enables a replay of any recording after the fact.
Replay runs proceed deterministically so that developers can harden
their code and check that the fixes remove the error.

\subsubsection{\FT{} Internals}

\FT{} takes advantage of Julia's operator overloading to track exceptions and
fuzz for vulnerabilities.
For example, below is a variant of \code{+} that is overloaded
for \code{TrackedFloat64}.
It calls the basic \code{+} (or injects a \Nan{}) and checks for exceptions
before returning:
This lets \FT{} intercept all \fp{} operations involving \texttt{TrackedFloat} types.

\begin{lstlisting}[language = Julia]
function +(x::TrackedFloat64, y::TrackedFloat64)
  result = run_or_inject(+, x.val, y.val)
  check_error(+, result, x.val, y.val)
  TrackedFloat64(result)
end
\end{lstlisting}

Every arithmetic operation requires a similar overloading.
Thus, the implementation of \FT{} uses a metaprogramming technique
adapted from \texttt{Sherlogs}~\cite{kMilanklSherlogsJl2021}
to abstract over the common patterns.
For every binary operation, the library creates an overloading similar
to the one for \code{+} above.
Unary operations and others work analogously.
This approach saved thousands of lines of code.
The implementation weighs in at 218 lines and defines 645 overloaded function;
assuming 5 lines of code per function, a handwritten version would require over 3,000 lines.

\subsection{Stack Graphs}
\label{s:cstg}

\FT{} can produce copious amounts of log files which can be challenging to sift through manually.
Coalesced stack trace graphs (CSTGs or \CSTG{}s for short) provide a way to visualize large amounts of stack traces in a compact form~\cite{humphreySystematicDebuggingMethods2014}.
This techinque pairs well with \FT{} and makes analyzing the log files easier.

\begin{figure}[t]
  \centering
  \includegraphics[width=0.9\columnwidth]{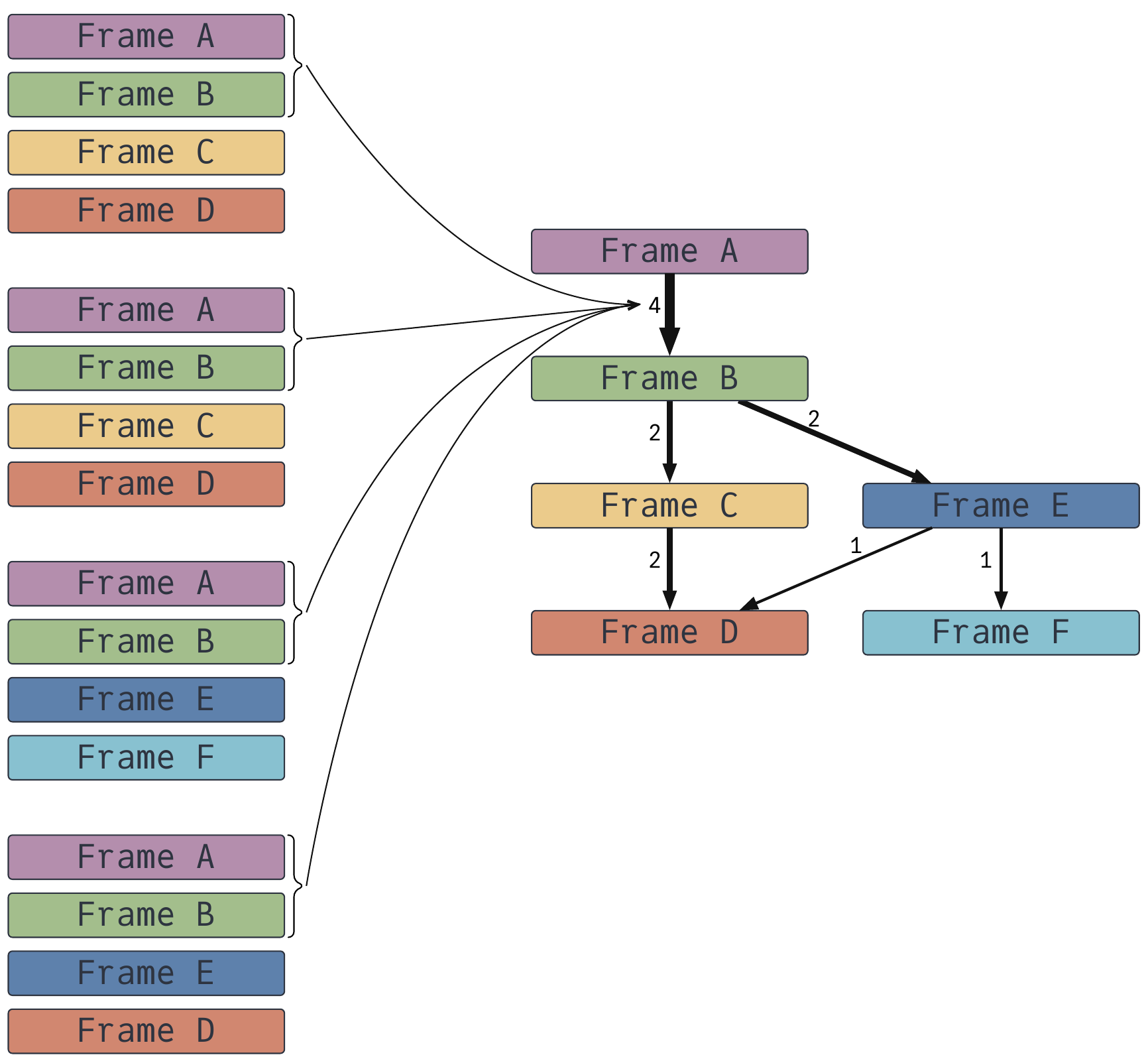}
  \caption{From stack traces (left) to \CSTG{} (right)}
  \label{fig:cstg_demo}
\end{figure}

\Cref{fig:cstg_demo} illustrates the construction of a \CSTG{} from a collection of stack traces.
Each trace on the left contributes nodes and edges to the graph on the right.
Repeated edges get emphasized with darker lines and larger counts.

Reading bottom-up, a \CSTG{} based on the \emph{gen} events in a program
highlights the contexts that frequently produced exceptional values.
Using~\cref{fig:cstg_demo} as an example, \code{Frame D} produced
three exceptions, two of which arose under \code{Frame C}.
In a large program with many exceptions, \CSTG{}s offer a way to prioritize
debugging efforts: go for the heavily-trodden paths first.

\subsection{\GPUFPX{}}
\label{s:gpufpx}

Many programs offload work onto GPUs, which are no less susceptible to \fp{} exceptions than CPUs.
In fact, GPU programs are worse off because they lack exception-handling mechanisms from the CPU world~\cite{llg-soap-2022}.
Since \FT{} instruments Julia programs, it cannot help directly;
however, the companion tool \GPUFPX{} instruments GPU kernels to detect and report
\fp{} exceptions~\cite{llsflg-hpdc-2023}.
Together, \FT{} and \GPUFPX{} provide insights for accelerated programs.

\section{Case Studies}
\label{s:casestudies}

\FlowFPX{} has helped to debug exceptions and fuzz for issues in a variety
of settings, some synthetic and some realistic.
The case studies include a shallow water simulation, the
\texttt{OrdinaryDiffEq} solver, and a Bayesian inference library.

\subsection{ShallowWaters}
\label{s:sw}

\texttt{ShallowWaters} simulates the flow of water over a
seabed~\cite{klowerNumberFormatsError2020,klowerPositsAlternativeFloats2019}.
The library has dozens of parameters that a scientist can experiment with.
One notable parameter is the Courant-Friedrichs-Lewy (CFL) number, which roughly describes the size of the time step to take in running the simulation.
A small CFL number makes the simulation run slowly, but accurately; a large number speeds it up but loses precision
because the system does not get enough time to propagate information.

\begin{figure}[t]
  \centering
  \includegraphics[width=3in]{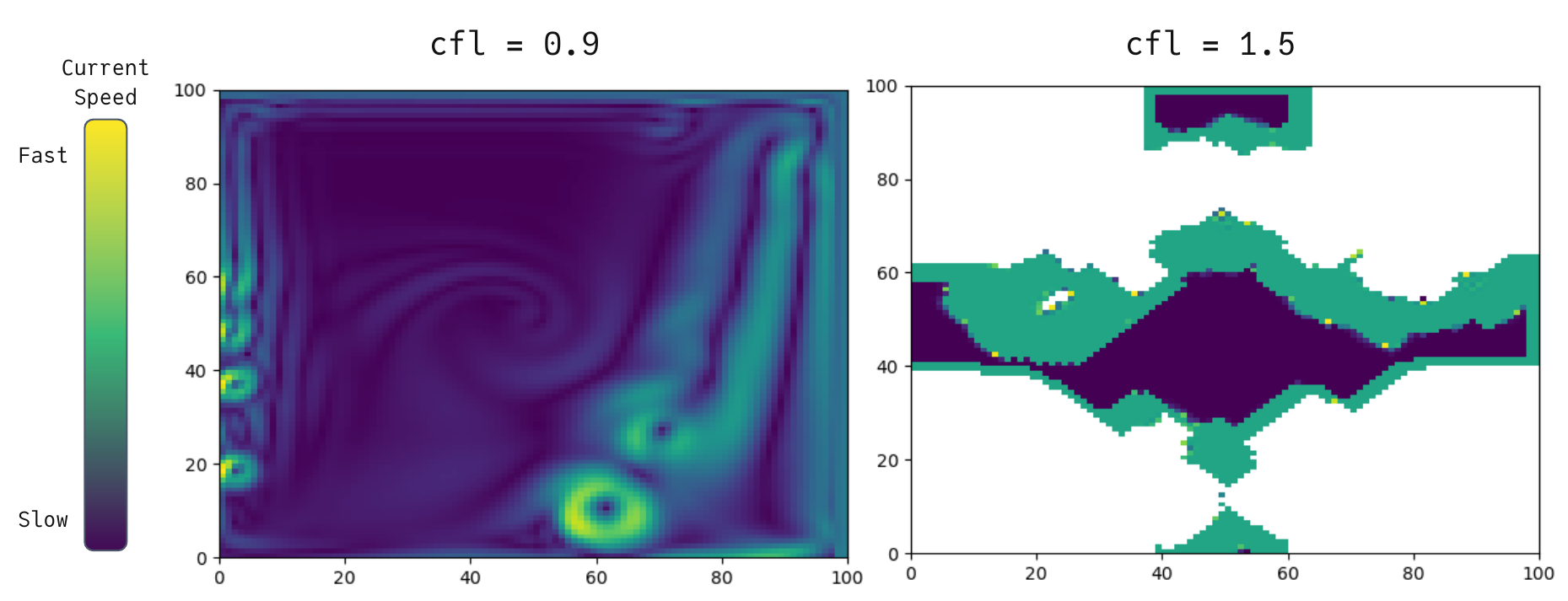}
  \caption{Raising the CFL number creates white gaps due to \NaN{}s}
  \label{fig:sw_nans}
\end{figure}

Normal values for the CFL number range between zero and one.
With a CFL of 0.9, the shallow water simulation produces the graph on the left half
of~\cref{fig:sw_nans} showing current speed in a rectangular basin.
Raising the CFL too high, to 1.6, causes trouble
(\cref{fig:sw_nans}, right) with large, white regions where the
current speed is \NaN{} instead of a normal value.

Running \FT{} on the high-CFL simulation reveals where the simulation drifted
from numbers to \NaN{} exceptions.
The first step in applying \FT{} is to convert relevant floats to tracked
floats, e.g., \code{Float32} to \code{TrackedFloat32}.
For \texttt{ShallowWaters}, this step is easy because the simulation is parameterized by
a \fp{} type for internal use.
Swapping in a tracked type is enough:

\begin{lstlisting}[language = Julia]
run_model(T=TrackedFloat32,
          cfl=10,
          nx=100,
          Ndays=100,
          L_ratio=1,
          bc="nonperiodic",
          wind_forcing_x="double_gyre",
          topography="seamount")
\end{lstlisting}

With tracking, the simulation runs as before, producing an ugly graph.
It additionally outputs logs for all gen, prop, and kill events
as they happen.
Below is one event from the gen logs:

\begin{lstlisting}
-([-Inf, -Inf])   FT/TrackedFloat.jl:106
momentum_u!       SW/rhs.jl:246
rhs_nonlinear!    SW/rhs.jl:50
rhs!              SW/rhs.jl:14 [inlined]
time_integration  SW/time_integration.jl:77
run_model         SW/run_model.jl:37
top-level
\end{lstlisting}

We can see that the \NaN{} appeared as the result of subtracting two infinities (\code{-Inf - -Inf}).
The trace further shows that this subtraction happens inside the function \texttt{momentum\_u!} on line 246 of the \texttt{rhs.jl} file.

This solves the mystery of where one \NaN{} came from, but raises a new question
about the source of the \Inf{} value.
The logs for \Inf{} gens have an answer:

\begin{lstlisting}
^([-1.515f31, 2])          FT/TrackedFloat.jl:138
literal_pow()              intfuncs.jl:325
...
materialize(^)             broadcast.jl:860
top-level getproperty(...) examples/sw_nan_tf.jl:14
..
\end{lstlisting}
This \Inf{} came from an exponent that overflowed the float type (\code{-1.515e31\string^2}).
\FT{} has shown exactly which numeric values in which operation
caused exceptions to occur.

\subsubsection{Stack Graphs for a Bigger Picture}

While the logs for \texttt{ShallowWaters} contain useful information,
there is an overwhelming amount of it.
There are over ten thousand lines in the gen file alone.
Converting these logs to a stack graph gives a quick overview
of the most common paths to exceptional values.

\begin{figure}[t]
  \centering
  \includegraphics[width=0.96\columnwidth]{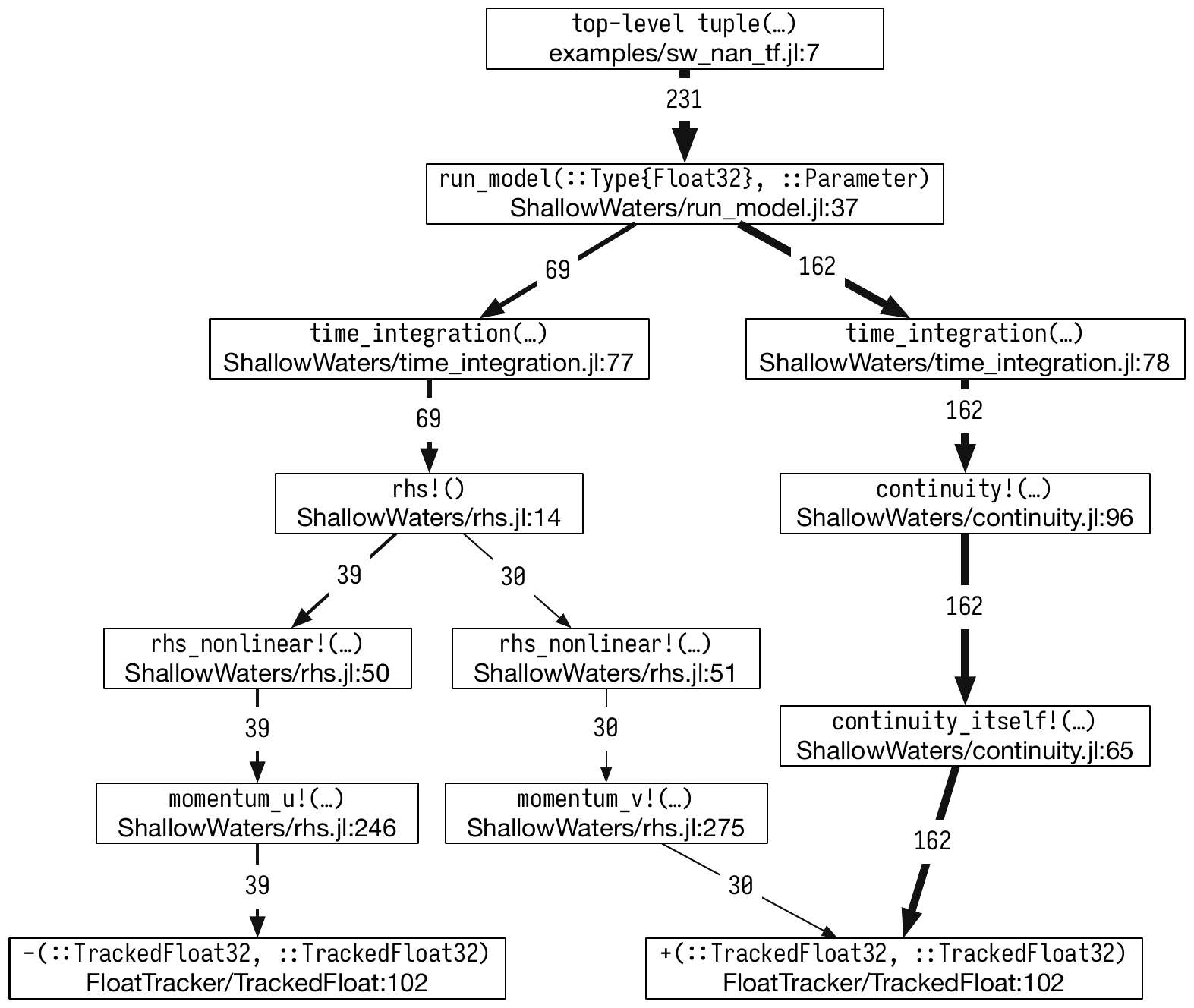}
  \caption{Stack graph for \NaN{} gens in \texttt{ShallowWaters}}
  \label{fig:sw_nan_cstg}
\end{figure}

\Cref{fig:sw_nan_cstg} presents the stack graph for \NaN{} gen events
in \texttt{ShallowWaters} with the high CFL number.
Reading bottom-up, every \NaN{} came from calls to the `\code{-}' and `\code{+}' operations.
Calls to `\code{+}' account for most of the \NaN{}s.
These \NaN{}s arose in two different contexts: a small number (30)
occurred within a momentum calculation, while the rest (162)
occurred within a continuity step.
Moving to the top of the graph, it shows that the function \code{run\_model} drove
the entire simulation.
With this overview of the program, a promising next step is to guard against \NaN{}s
in the momentum and continuity functions.

% \begin{figure}[t]
%   \centering
%   \includegraphics[width=3in]{fig/sw_inf_cstg_clean.png}
%   \caption{CSTG of \Inf{} generation logs from the same run as~\cref{fig:sw_nan_cstg}.}
%   \label{fig:sw_inf_cstg}
% \end{figure}

\subsubsection{Stack Graph Differences}

\begin{figure}[t]
  \centering
  \includegraphics[width=0.96\columnwidth]{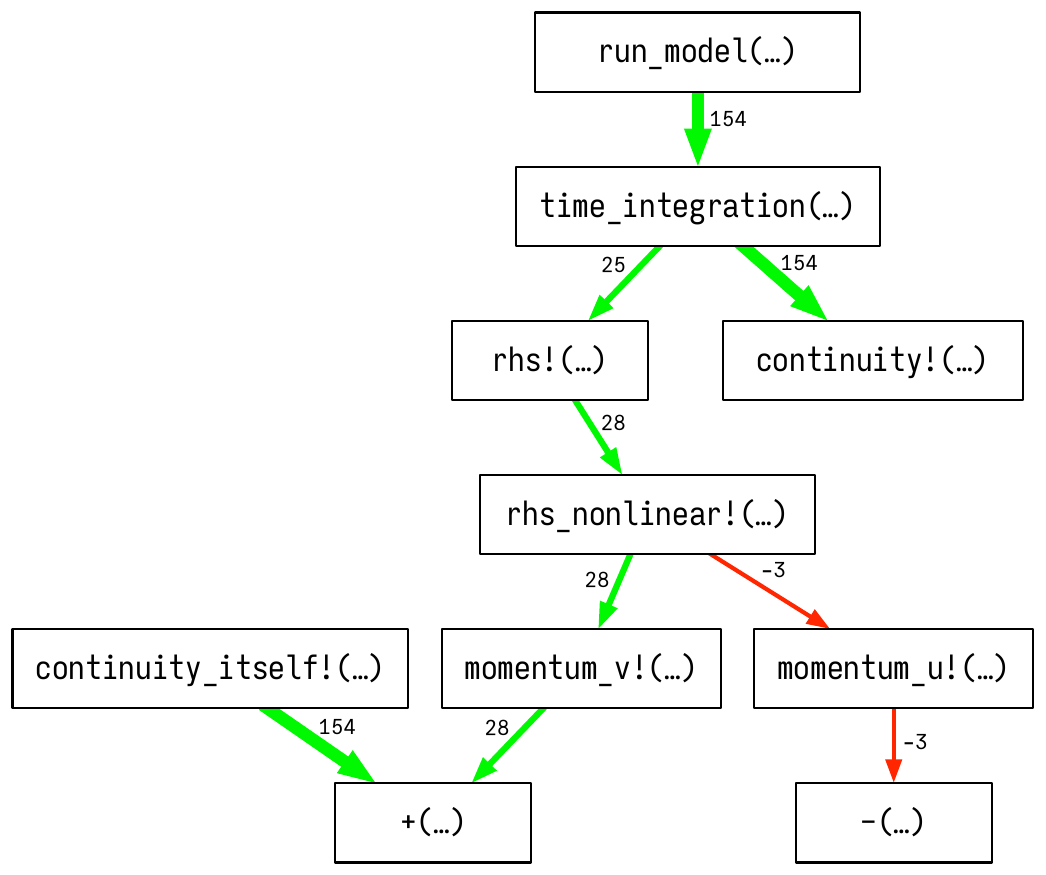}
  \caption{Stack graph diff}
  \label{fig:cstg_diff_demo}
\end{figure}

Graph diffing works well for \CSTG{}s; it shows how flows in the program evolved
from one stage to another.
\Cref{fig:cstg_diff_demo} illustrates one diff in the context of \NaN{} gens for
\texttt{ShallowWaters}:
it compares the first 10\% of gen events to the latter 90\% of gens.
The positive numbers and green lines indicate flows that are new in the latter part,
and the negative numbers and red lines show flows that disappeared in the latter part.
A domain expert might use these clues to find where an instability started in
the first part of the program.
In this case, the function \texttt{momentum\_u!} showed up only in the
beginning of the logs---it may be an effective point to check for \Nan{}s.

\subsection{OrdinaryDiffEq}
\label{s:ode}

With the targeted fuzzing abilities of \FT{}, we set focus on the
\texttt{NBodySimulator} package from the SciML
team.\footnote{\url{https://github.com/SciML/NBodySimulator.jl}}
We only wanted to test the robustness of the \texttt{NBodySimulator} package, so we configured the fuzzer to inject \NaN{}s \emph{only} when inside of a function of that library.
This way, we could avoid injecting in the standard library or other dependencies of \texttt{NBodySimulator} and concentrate on improving this one library.

However, Fuzzing discovered zero bugs and zero exceptional events.
In fact, fuzzing injected \emph{zero} exceptional values whatsoever, even with adjusting the odds to \emph{force} an injection at the first available opportunity.
The reason for this lack of \NaN{} injection, it turns out, is that the nbody
simulation merely creates a problem for the \code{OrdinaryDiffEq}
solver.\footnote{\url{https://github.com/SciML/OrdinaryDiffEq.jl}}
All the \fp{} operations were in another library.

Fuzzing on \code{OrdinaryDiffEq} led to a curious situation.
The library itself reported a \NaN{} and printed a message that it would exit.
However, after printing that message, the program went into an infinite loop.
Using \CSTG{}s to guide our search, we found a \NaN{} kill that manifested repeatedly in the logs.
The stack traces for that kill originated from inside the file \texttt{solve.jl}:

\begin{lstlisting}
<([NaN, 3.0e6])  at FT/TrackedFloat.jl:193
solve!           at ODE/solve.jl:515
...
\end{lstlisting}

The relevant part of \texttt{solve.jl} contains a pair of loops.
With injection, the variable \code{tdir} holds a \NaN{},
stopping all productivity:

% file here: ~/Research/ode_debug/dev/OrdinaryDiffEq/src/solve.jl
% see line 514

\begin{lstlisting}[language = Julia]
while !isempty(time_stops)
  while tdir * t < first(time_stops)
    # do integration work
    # pop_off(time_stops)
  end
end
\end{lstlisting}

In more detail, the \NaN{} for \code{tdir} propagates though the multiplication
and gets killed by the \code{<} comparison.
Hence the condition for inner \code{while} loop is always false,
which means the outer loop never ends up with an empty list.

This is a real-world example of how \NaN{} kills can affect control flow, and we filed
an issue for it.\footnote{\url{https://github.com/SciML/OrdinaryDiffEq.jl/issues/1939}}
Fortunately, the problem in this case is benign as the code was already trying to halt.

\subsection{Finch}
\label{s:finch}

Finch is a domain-specific language for specifying
PDEs~\cite{heislerFinchDomainSpecific2022}.\footnote{Not to be confused
with the Finch loop optimizer~\cite{adka-cgo-2023}.}
In the spirit of FEniCS~\cite{fenics} and related
tools~\cite{freefem,openfoam,dune,firedrake},
Finch helps scientists quickly convert math into code.
What sets Finch apart is its flexibility.
It supports multiple discretization methods (finite element and finite
volume) and multiple backends (Julia, \CPP{}, \Dendro{}~\cite{dendro}).
Furthermore it strives to output code that humans can easily fine-tune.

Fuzzing with \FT{} revealed two places where Finch needed
protection against user input.
The first was when reading an input mesh.\footnote{\urlaccess{https://github.com/paralab/Finch/issues/16}{2023-06-06}}
A \NaN{} injected in the mesh led to a crash further on:

\begin{lstlisting}
  BoundsError: attempt to access 1-element
   Vector{Int64} at index [2]
\end{lstlisting}

Before accessing the input, Finch needed to check for \NaN{}s.
The second place was in setting bounds for the
solver.\footnote{\urlaccess{https://github.com/paralab/Finch/issues/17}{2023-06-06}}
Here, a \NaN{} could leave bounds uninitialized, leading to a bounds error.
Additionally, \FT{} and \CSTG{}s have been useful for identifying \NaN{}s that
appear in unstable heat simulations written in Finch.
%% advection2d, high cfl and high T (time) led to NaN during simulation

\subsection{Oceananigans}
\label{s:ocean}

\texttt{Oceananigans}~\cite{OceananigansJOSS} is simulation package for incompressible
fluid dynamics that can generate code for Nvidia GPUs.
For example, the following program (from the project readme) simulates turbulance:

\begin{lstlisting}[language = Julia]
using Oceananigans
grid = RectilinearGrid(GPU(),
  size=(128, 128), x=(0, 2π), y=(0, 2π),
  topology=(Periodic, Periodic, Flat))
model = NonhydrostaticModel(; grid,
  advection=WENO())
ϵ(x, y, z) = 2rand() - 1
set!(model, u=ϵ, v=ϵ)
simulation = Simulation(model;
  Δt=0.01, stop_time=4)
run!(simulation)
\end{lstlisting}

\GPUFPX{} provides detailed feedback on this program.
The output in~\cref{f:gpufpx} shows that 21 kernels
appear and generate six floating-point exceptions.
There are three \NaN{}s, one \Inf{}, and two division
by zero errors.
The report is a starting point for further investigation
of the reliability of the example.
%% TODO exactly how?

\begin{figure}[t]\centering
  \begin{tabular}[t]{ll}
    \begin{tabular}[t]{lr}
      \zerocode{-{}- FP64 Operations -{}-} \\
      \code{Total NaN:} & \code{2} \\
      \code{Total INF:} & \code{1} \\
      \code{Total subnormal:} & \code{0} \\
      \code{Total div0:} & \code{2} \\
    \end{tabular}
    &
    \begin{tabular}[t]{lr}
      \zerocode{-{}- FP32 Operations -{}-} \\
      \code{Total NaN:} & \code{1} \\
      \code{Total INF:} & \code{0} \\
      \code{Total subnormal:} & \code{0} \\
      \code{Total div0:} & \code{0} \\
    \end{tabular}
  \end{tabular}
  \begin{tabular}{lr}
    \zerocode{-{}- Other Stats -{}-} \\
    \code{Kernels:} & \code{21}
  \end{tabular}

  \caption{Example \GPUFPX{} output}
  \label{f:gpufpx}
\end{figure}

\subsection{RxInfer}
\label{s:safari}

We discovered an open issue in the \code{RxInfer.jl} library related to \NaN{} detection
and suggested \FT{} to the developers.\footnote{\url{https://github.com/biaslab/RxInfer.jl/issues/116}}
Within a day, they were able to track down the location of an important \NaN{} gen.
This success came with little input from our end; in fact, the application program was
closed-source.
We merely explained how to use \FT{} by wrapping inputs in a tracked type.

\section{Related Work}
\label{s:related}

\subsection{Error Analysis}

Demmel et al.\cite{ddghlllprr-correctness-2022} examine \fp{} exceptional value handling in BLAS and LAPACK, identify several inconsistencies, and propose an API for debugging and adding determinism.
The proposal includes an extension to the \texttt{INFO\_ARRAY} parameter with fields that record \genpropkill{} information.
It also includes fuzzing, which we realize in \FT{}.

Toronto and McCarthy~\cite{torontoPracticallyAccurateFloatingPoint2014} propose a test-driven method for detecting numeric error: plot the results of an expression on a range of inputs and look for sharp deviations, or \emph{badlands}.
They point out several ways to rewrite code to avoid badlands by rewriting code in a semantics-preserving way.

Herbie~\cite{panchekhaAutomaticallyImprovingAccuracy2015} automatically rewrites arithmetic expressions to reduce \fp{} error.
This would combine well with \FT{}: first identify the location of a \NaN{}, ask Herbie to find a repair.
The Odyssey\cite{misbackOdysseyInteractiveWorkbench2023} workbench provides an interactive interface to Herbie.

\subsection{Diagnosing floating-point exceptions}

The Julia library \texttt{Sherlogs}~\cite{kMilanklSherlogsJl2021} inspired our use of a custom number type to intercept operations on a number.
In contrast to \FT{} which monitors for exceptional values and logs stack traces at interesting points in their lifetime, \texttt{Sherlogs} tracks and reports the range of values seen over the course of a computation.
This is intended to provide insight into whether or not a library could tolerate a lower-precision \fp{} format.

% TODO: make sure that what I said about what FPSpy can track is true.
FPSpy~\cite{dindaSpyingFloatingPoint2020} is an \texttt{LD\_PRELOAD} shared library that works on unmodified x86 binaries.
It monitors a program during execution for operations that generate an exception, such as division by zero, underflow, and overflow.
By contrast to \FT{}, it does not track prop or kill events.
FPSpy has the advantage of being lightweight enough to run on production code for certain loads.

\subsection{Stack Graphs}

Our stack graphs utilize the CSTG library for coalesced stack trace graphs~\cite{humphreySystematicDebuggingMethods2014}.
In turn, CSTGs build on the STAT tool from LLNL~\cite{arnoldStackTraceAnalysis2007}.
STAT collects, analyzes, and visualizes stack traces from concurrent processes
to highlight anomalies.
It produces visualizations similar to those of CSTG, thought CSTG offers more compact views and supports diffs.

\subsection{GPU Exception Tracking}

FPChecker~\cite{l-ase-2019} is a tool to report \fp{} exceptions occurring on the GPU.
FPChecker relies on LLVM-level instrumentation of GPU kernels, and so cannot run on the plethora of closed-source GPU kernels in usage today.
BinFPE~\cite{llg-soap-2022} is another tool in the same space;
BinFPE performs SASS-level analysis of GPU kernels, but is limited in that it is slow and does not catch errors that alter control flow.
The latter deficiency is particularly worrisome, as we have seen, silent \NaN{} kills can invalidate results without the user noticing.
\GPUFPX{}~\cite{llsflg-hpdc-2023} improves on the work of FPChecker and BinFPE by being more performant and catching a wider set of errors, including those that alter control flow.
\GPUFPX{} is a shared library that, like FPSpy uses \texttt{LD\_PRELOAD} to work on unmodified binaries.
\GPUFPX{} runs on CUDA cores from NVIDIA and reports total numbers of exceptional values.
Like \FT{}, \GPUFPX{} can catch \NaN{} gens and kills, but it does this on the GPU where \FT{} doesn't apply.
Despite these improvements, \GPUFPX{} is limited by the closed-source nature of common GPU cores, and cannot report at the rich level of source detail that \FT{} can.

% % Herbie, improving accuracy automatically.
% % ParetoHerbie

% %% http://eprints.maths.manchester.ac.uk/2873/1/fami22.pdf
% %% https://drive.google.com/drive/folders/1HkgMpxi6hsqLSZj9tSnuCxnFExPPjmvx?usp=sharing
% %% https://docs.google.com/presentation/d/1ueqzb5QY9YHmtY2DfY_hpN9oufSOcXErLWsJOZQxgzo/edit?usp=sharing
% %% https://ieeexplore.ieee.org/document/8916392
% %% https://link.springer.com/chapter/10.1007/978-3-319-73721-8_24

% FPChecker, " considers CPU OpenMP and MPI codes, again using LLVM
% instrumentation"~\cite{ltlg-iiswc-2022}
% % FPChecker has a nice table of tools; some already mentioned here

% 6-8,24,25
% Arnab Das scalable yet rigorous
% Marc Daumas Certification of Bounds Expres
% David Delmas Towards an Industrial Use of
% Alexey Solovyev FPTaylor Results table
% Titolo absint

% GPU-FPX state of art, companion project~\cite{llsflg-hpdc-2023}.
% Latest in line of work.
% BinFPE exn checking, slower than GPU-FPX, 
% \cite{llg-soap-2022}.
% FPChecker, early pioneer, fpx in gpus, llvm-level instrumentation
% if compiled with Clang, but most gpus use closed NVCC compiler~\cite{l-ase-2019}.

\section{Discussion}
\label{s:discussion}
% future work

Lightweight tools for error analysis that can quickly identify
\fp{} problems and suggest repairs are an important topic.
The number of scale of scientific application has grown tremendously
over the years.
For small teams that cannot afford a full-time
analyst, tools like \FlowFPX{} fill a critical role.

\FlowFPX{} it itself an evolving toolkit.
Below we discuss some topics for future work.

\subsection{Performance}
\label{s:discussion-performance}

\FT{} incurs significant overhead on the order of 100x slower than a non-instrumented run of the same program.
It is a debugging tool, not a production tool.
To put this number into context, Valgrind runs with a similar level of slowdown.

In addition to the cost incurred by intercepting \fp{} operations, gathering stack traces is expensive.
We observe a 10x slowdown on \texttt{ShallowWaters} with logging disabled.
We recommend that users of \FT{} make use of the \texttt{maxLogs} and \texttt{exclude\_stacktrace} configuration options to limit the number and kind of logs gathered, and thereby reduce the number of calls to \texttt{stacktrace()}.
Stack traces are essential to decide where to inject a fault when fuzzing,
but we defer them as late as possible to maximize performance.

\subsection{Enhanced Fuzzing}

While fuzzing is useful for discovering issues, its success rate
is low because {every} \fp{} operation is a candidate
for injection.
Even operations that are already well-defended against \NaN{}s are candidates.
\FT{} could use two sorts of tools for improving injection.
First, fine-grained control to let users decide where not to inject.
Second, tools for understanding the context of an injection point
after the fact.
Program slicing is especially relevant to the latter point and effectively
what we did by hand when fuzzing Finch (\cref{s:finch}).
For each operation, an expert needs to study the values that feed into it to
decide whether they are protected or not.

%% TODO example, start here at glbvertex https://github.com/paralab/Finch/blob/master/src/grid.jl#L360

\subsection{Tracking Exceptions in External Libraries}

\FT{} is limited to Julia code.
It cannot track the lifetime of exceptional values in external libraries, such as GPU kernels or C programs.
For GPUs it relies on \GPUFPX{}, through the connection between these tools is loose.
In the future, \FT{} would benefit from an API to plug in tools for external libraries, gather their output,
and present a comprehensive view of exceptions in a multi-language program.

\subsection{Interface Concerns}

FloatTracker is capable of monitoring all sorts of events and \fp{} values beyond exceptions.
For example, \FT{} could monitor large or small values with guidance from the user about which values are worth logging.
On a similar note, the interface to \FT{} is primitive: users express interest in a number by wrapping it
in a constructor such as \code{TrackedFloat64}.
Helper functions that track data structures and choose a default size would
make it easier to experiment with \FT{}.

\section{Acknowledgments}

Thanks to the \code{Sherlogs.jl} developers for inspiring the architecture of \FT{}.
Thanks to the \code{RxInfer.jl} developers for testing \FT{} and for feedback on its user interface.
Thanks to Alex Larsen and Rob Durst for comments on an early draft.
This work is supported by
NSF grant \href{https://nsf.gov/awardsearch/showAward?AWD_ID=2030859&HistoricalAwards=false}{2030859}
to the CRA for the \href{https://cifellows2020.org}{CIFellows} project,
and DOE ASCR Award Number DE-SC0022252 and NSF CISE Awards 1956106 and 21241.

\input{bib.tex}

\end{document}

%% file: header.tex
% **************GENERATED FILE, DO NOT EDIT**************

\title{FlowFPX: Nimble Tools for Debugging Floating-Point Exceptions}

\author[1]{Taylor Allred}
\author[1]{Xinyi Li}
\author[1]{Ashton Wiersdorf}
\author[1]{Ben Greenman}
\author[1]{Ganesh Gopalakrishnan}
\affil[1]{University of Utah}


\hypersetup{
pdftitle = {FlowFPX: Nimble Tools for Debugging Floating-Point Exceptions},
pdfsubject = {JuliaCon 2022 Proceedings},
pdfauthor = {Taylor Allred, Xinyi Li, Ashton Wiersdorf, Ben Greenman, Ganesh Gopalakrishnan},
pdfkeywords = {Julia, floating-point, debugging},
}

%% file: bib.tex
% **************GENERATED FILE, DO NOT EDIT**************

\bibliographystyle{juliacon}
\bibliography{ref.bib}